\documentclass[apj, chicago]{emulateapj}
\journalinfo{\textsc{The Astrophysical Journal}}
\slugcomment{Received 27 November, 2018; Accepted 19 February 2019}
\usepackage{url}

\usepackage{natbib}
\usepackage{epstopdf}
\usepackage{graphicx}
\usepackage[colorlinks,
        hypertexnames=false,
        pagebackref=true,
        linkcolor=blue,    
    citecolor=blue,        
    filecolor=magenta,     
    urlcolor=blue          
]{hyperref}                

\usepackage{array,etoolbox}
\preto\tabular{\setcounter{magicrownumbers}{0}}
\newcounter{magicrownumbers}

\shorttitle{A solar eruption from a weak field region}
\shortauthors{Vasantharaju et al}

\begin{document}
\title{Formation and eruption of sigmoidal structure from a weak field region of NOAA 11942}
\author{N.~Vasantharaju$^1$, P.~Vemareddy$^2$, B.~Ravindra$^2$, and V.~H.~Doddamani$^1$}
\email{vrajuap@gmail.com}
\affil{$^1$Department of Physics, Bangalore University, Bengaluru-560 056, India}
\affil{$^2$Indian Institute of Astrophysics, Koramangala, Bengaluru-560 034, India}

\begin{abstract}
Using observations from Solar Dynamics Observatory, we studied an interesting example of a  sigmoid formation and eruption from small-scale flux canceling regions of active region (AR) 11942. Analysis of HMI and AIA observations infer that initially the AR is compact and bipolar in nature, evolved to sheared configuration consisting of inverse J-shaped loops hosting a filament channel over a couple of days. By tracking the photospheric magnetic features, shearing and converging motions are observed to play a prime role in the development of S-shaped loops and further flux cancellation leads to tether-cutting reconnection of J-loops. This phase is co-temporal with the filament rise motion followed by sigmoid eruption at 21:32 UT on January 6. The flux rope rises in phases of slow  (v$_{avg}$ = 26 km~s$^{-1}$) and fast  (a$_{avg}$= 55 ms$^{-2}$) rise motion categorizing the CME as slow with an associated weak C1.0 class X-ray flare. The flare ribbon separation velocity peaks at around peak time of the flare at which maximum reconnection rate (2.14 Vcm$^{-1}$) occurs. Further, the EUV light-curves of 131, 171\AA~have delayed peaks of 130 minutes compared to 94\AA~and is explained by differential emission measure. Our analysis suggests that the energy release is proceeded in a much long time duration, manifesting the onset of filament rise and eventual eruption driven by converging and canceling flux in the photosphere. Unlike strong eruption events, the observed slow CME and weak flare are indications of slow runway tether-cutting reconnection where most of the sheared arcade is relaxed during the extended post phase of the eruption.
\end{abstract}

\keywords{Sun:  reconnection--- Sun: flares --- Sun: Sigmoid ---Sun: coronal mass ejection --- Sun: magnetic fields---Sun: non-potentiality}
\section{Introduction}
\label{Intro}
Solar coronal mass ejection (CME) has been a major objective of research for the past two decades due to its ability to affect the space weather phenomena tremendously. Hence, many studies have been conducted towards forecasting such eruptions by finding the suitable eruptive conditions from its source regions \citep{Falconer2001,Falconer2003,Leka2003a,Leka2003b,Schrijver2007,Vemareddy2015,Vasantharaju2018}. Also, the appearance of S-shaped structure or inverse S-shaped structure or simply ``Sigmoid" in the coronal active region is considered to be a progenitor of CMEs \citep{Rust1996,Chen2011}. Generally the sheared and twisted field lines in sigmoidal structure possesses large amount of magnetic free-energy and will be released during CME. Mostly, S-shaped sigmoids are observed in the southern hemisphere and inverse-S shaped ones are observed in the northern hemisphere \citep{Rust1996,Pevtsov2001}. This is consistent with the predominantly negative current helicity observed  in the northern hemisphere and positive in the southern hemisphere \citep{Seehafer1990,Pevtsov1995}. Active regions possessing sigmoids are significantly more likely to be eruptive than non-sigmoid active regions \citep{Canfield1999}. The sigmoid is now regarded as an important signature in space weather forecasts \citep{Rust2005,Vemareddy2015}.

Past statistical studies confirm that majority of CMEs have a clear magnetic flux rope (MFR) structure \citep{Vourlidas2013}. Since the coronal magnetic field cannot be measured directly, the flux rope model can only seek for indirect evidence from observations.  \citet{Rust1996} proposed that sigmoidal structure in AR can be explained by the flux rope model. Later several studies explored the flux rope model with numerical experiments which confirm by reproducing many of the observational properties of the sigmoid \citep{Gibson2006}.

The major questions about these flux ropes are how they are formed and how they are initiated to produce an eruption. Sigmoid eruptions are interpreted by two kinds of observational models based on two different view points of the same structure as sheared arcade and flux rope.  The first kind of models assume that the sigmoid is composed of sheared and twisted core field in the AR and the internal/external runaway tether‐cutting reconnection is responsible for sigmoid to arcade transformation and eruption \citep{Antiochos1999,Moore2001}. On the other hand, the second kind of models argue that the evolution of a sigmoidal AR with flare and CME is explained by a twisted magnetic flux rope in the form of filament structure that emerges and equilibrates with the overlying coronal magnetic field structure \citep{Gibson2006}. \citet{VanBallegooijen1989} proposed a scenario of flux rope formation from a sheared arcade where flux cancellation occurs by slow shearing and converging regions about the polarity inversion line (PIL).   
	
From the magnetohydrodynamic (MHD) point of view, the flux rope is in equilibrium under the balance of magnetic pressure in the flux rope and the magnetic tension of the overlying magnetic field. If the twist number increases to some critical value, then kink instability would occur \citep{Hood1979,Torok2004}. If the decay index of the background field, in which the MFR is embedded, is larger than some critical value, torus instability can occur \citep{Kliem2006,Aulanier2010}. However, in general both ideal MHD instability and magnetic reconnection mechanisms are responsible for release of magnetic energy by triggering the flux rope eruption successfully \citep{Forbes2000}. These models are similar to catastrophe model where the flux rope would lose equilibrium after reaching a critical height, forming a current sheet beneath it, connecting the post-CME loops to the CME ejecta \citep{Lin2005,Bemporad2006}, as predicted by both the classical CSHKP flare model \citep{Carmichael1964,Sturrock1966,Hirayama1974,Kopp1976} and \textbf{other} CME models (e.g., \citealt{Lin2000,Chen2007}). 

Generally,  the catastrophe model explain efficiently the observed CME-flare associated eruptions rather than flare-less CME events. Quite recently \citet{Song2013} studied the energy release mechanisms in a sample of 13 flare-less CME events, which are associated with prominence eruptions originated from relatively weak field quiet-Sun regions. They found that the ideal MHD flux-rope instability plays a major role in magnetic energy release process while the magnetic reconnection plays the minor role. This manuscript presents a comprehensive study of formation, initiation and eruption of a sigmoid from a weak field AR. We try to understand the weak eruption via probing questions like 1) Under what conditions it became eruptive?, 2) what conditions made it weak (in terms of CME speed, energy and associated flare class)?, and 3) How are the observed consequences different from strong eruption events?. In section~\ref{ObsData}, we described the observational data. The results are presented in section~\ref{Res} and summarized with conclusion in section~\ref{summ}. 

\section{Observational Data}
\label{ObsData}
The Atmospheric Imaging Assembly (AIA; \citet{Lemen2012}) on-board Solar Dynamics Observatory (SDO; \citet{Pesnell2012}) produces full-disc Extreme ultraviolet (EUV) images in 10 wavelength bands at a high cadence of 12 seconds and pixel size of 0\arcsec.6. For the present study, we used the images obtained in 94\AA (Fe XVIII; T$ \sim $6.4MK), 131\AA (Fe XXI ;T$ \sim $10MK), 171\AA(Fe IX;T~$ \sim $0.6MK), 193 \AA( Fe XII; T$ \sim $1.6MK \& Fe XXIV ;T$ \sim $20MK) and  304\AA ( He-II;T$ \sim $0.05MK) wavelengths. To increase the contrast and signal-to-noise ratio, we added five consecutive images to give the cadence of 1-minute. The corresponding photospheric magnetic field observations are obtained from the Helioseismic and Magnetic Imager (HMI; \citet{Scherrer2012}) on board SDO. Both line-of-sight(LOS) and vector magnetic field measurements obtained at a cadence of 12 minutes are used in this study. Vector data set is \texttt{hmi.sharp\_cea\_720s} series, which provides automatically selected cutouts of active regions (space weather HMI active region patches; SHARPs $ - $ see \citet{Bobra2014}) in the form of Lambert cylindrical equal-area (CEA) projection. Using the cut-out image processing aide provided by JSOC Stanford website, the cut-outs of  LOS full-disk magnetogram data were used and then these cut-outs are corrected for the area foreshortening that occurs away from central meridian. The CME was observed in white-light by the Large Angle and Spectrometric Coronagraph (LASCO; \citet{Brueckner1995}) on board SOHO, which consists of two optical systems, C2 (2.2 $-$ 6.0 $R_\odot$) and C3 (4 $ - $ 32 $R_\odot$).

 \begin{figure*}[!ht]
	\centering
	\includegraphics[width=.99\textwidth,clip=]{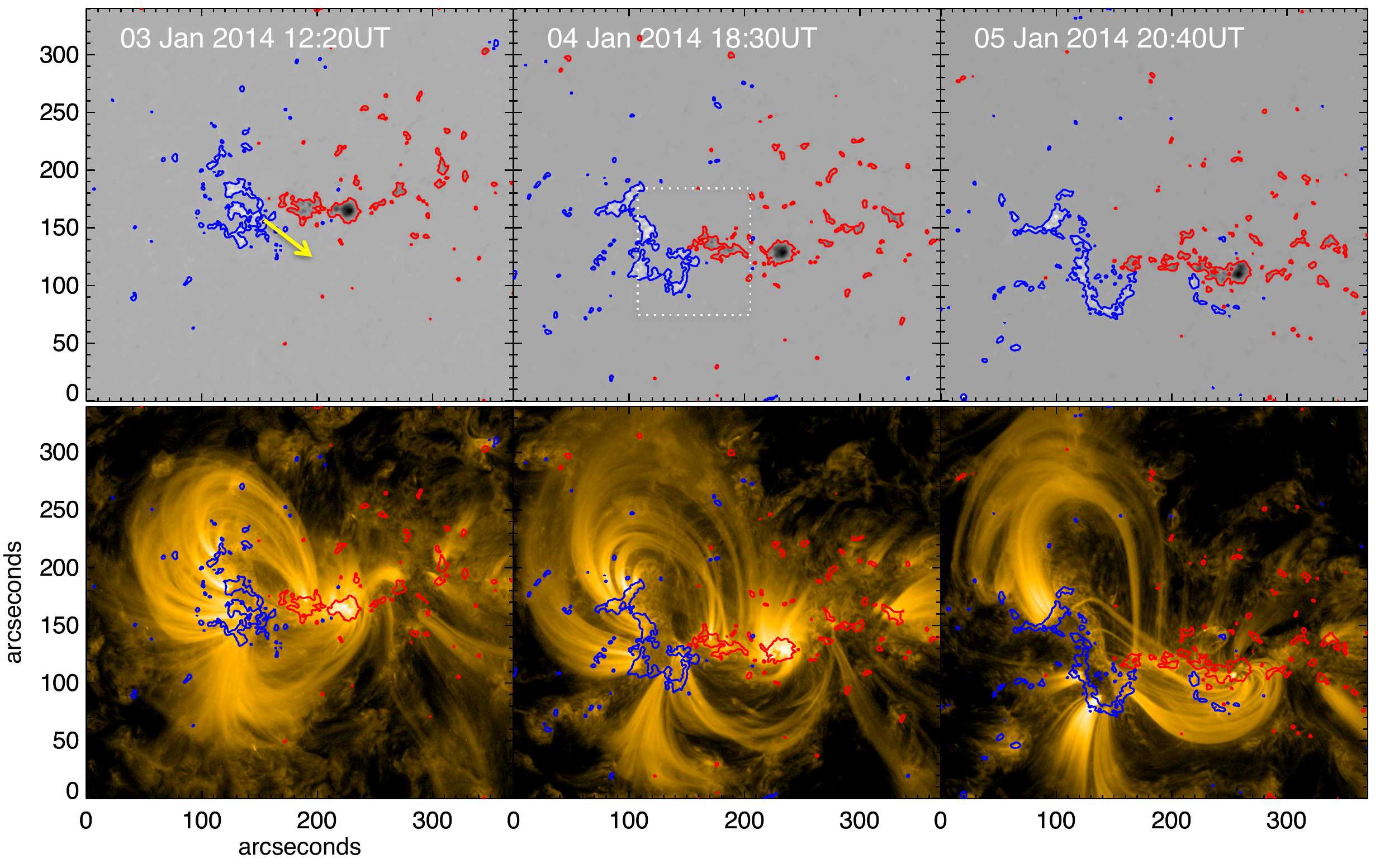}
	\caption{Observations of AR 11942 evolution over days. Top row: line-of-sight magnetic field observations with contours at $\pm$100G. Yellow arrow indicates the direction of shear motion of positive polarity and the dotted rectangular box in 04/18:30UT panel contains the interacting region of positive and negative polarities to further examine the flux motions.  Bottom row: AIA 171\AA~ observations showing build-up of inverse-S shape sheared arcade from potential like loop structure on January 3. Bz-contours at $\pm$100G are overlaid to identify the polarity of loop foot points.} 
	\label{Fig_overall}
\end{figure*}

\section{Analysis and Results}
\label{Res}
Active region NOAA 11942 first appeared as the compact region on January 1, 2014 in the eastern limb of northern hemisphere. It was of Mount Wilson class-$\alpha\gamma$ gradually evolving to class-$\beta\gamma$ while moving across the disk towards the west-limb and finally revert to class-$\alpha\gamma$ before it decays on 10 January 2014. During the disk passage, AR 11942  possessed an inverse-S sigmoidal structure in EUV wavebands during its decay phase on 6-7th January 2014. The fragmentation and dispersion of both north and south polarity regions lead to the formation of sigmoid and it supported the filament at initial stages. In due course the filament got separated from sigmoid and started to rise upwards leading to a CME with association of small X-ray flare. The inverse-S shape indicates the region has left-handed chirality or negative helicity \citep{Pevtsov2002}. Motivated by these observations, we studied the formation of sigmoidal structure, driving and triggering mechanisms of the eruption from sigmoidal AR.

\subsubsection{Slow evolution and build-up of sigmoidal loop structure}
\begin{figure*}[!ht]
	\centering
	\includegraphics[width=.85\textwidth,clip=]{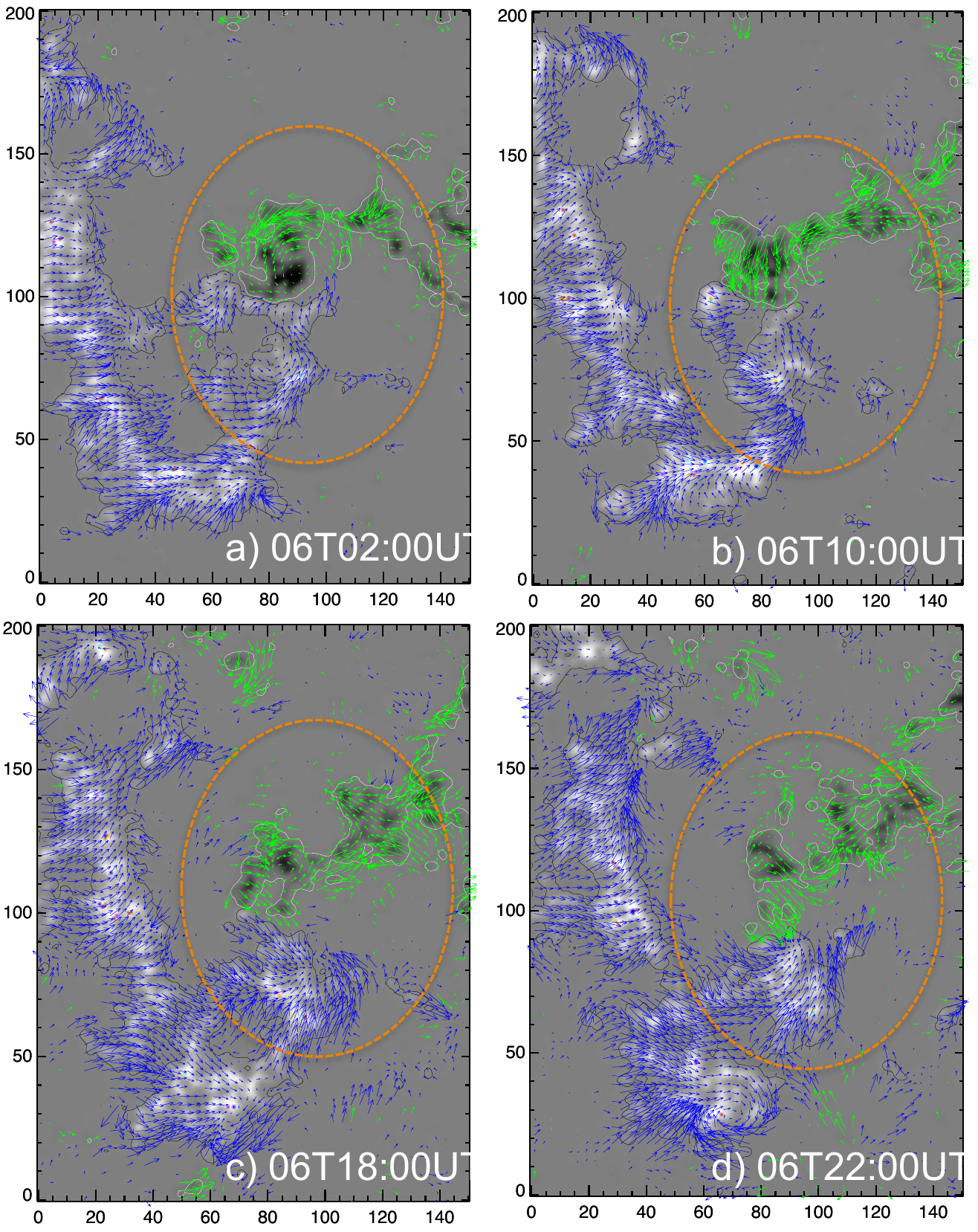}
	\caption{Horizontal velocity of flux motions in the interacting region of positive and negative polarities (rectangular box in Figure~\ref{Fig_overall}). The derived velocities from HMI LOS $B_z$ (12 minute cadence) are averaged over 4 hours to enhance the large scale motion of the fluxes. The arrows indicate direction of motion and their length is proportional to magnitude to a maximum of 0.6 km/s. Both positive and negative polarities in each panel move towards each other and interact by converging motion in a region enclosed by orange oval. In all panels, axis units are in pixels of 0.5\arcsec.} 
	\label{Fig_vel}
\end{figure*}

 To present the typical evolution of AR 11942, we used characteristic snapshots of AIA 171~\AA~ on different days overlaid with corresponding LOS magnetogram contours, shown in the bottom panels of Figure~\ref{Fig_overall}. The magnetograms are shown in the top panels. The AR consists of leading negative and following positive polarity mimicking a bipolar configuration. We noticed from these panels that the positive and negative polarity regions were in compact and less dispersed phase on January 3, 2014, and the coronal loop structure in AIA 171~\AA~images indicates a simple potential field configuration (first column of Figure~\ref{Fig_overall}). Further on January 4 (second column of Figure~\ref{Fig_overall}), the positive polarity regions start to diffuse towards the southern direction whereas negative polarity elements start to undergo further fragmentation and dispersion towards west. This process as a whole manifests converging and shearing motion of magnetic patches in the interfering region of the leading negative and following positive polarities (See also Figure~\ref{Fig_vel}). While this process continues further  till January 5 (third column of Figure~\ref{Fig_overall}), the AR loops became increasingly sheared with a morphology of inverse-S shape as being the combination of two inverse-J shaped loops joining at the middle. During this process, we also noticed emergence of positive magnetic flux near the negative polarity region towards west which further complicates the magnetic topology of the AR. We can see the difference of simple potential configuration on January 3 and that of inverse S-shape sigmoidal loop structure on January 5. 
 
\subsubsection{Shearing and Converging Motions}
\label{subsec_shear_fc}
Further to study the converging and shearing flux motions, we derived horizontal velocity field by \textit{differential affine velocity estimator} (DAVE; \citealt{Schuck2006}) technique. We use HMI LOS magnetogram maps at 12 minute cadence. In order to highlight large-scale and long term flow pattern, the velocity maps are averaged over 4 hours. These flow velocities are shown in Figure~\ref{Fig_vel}. Vectors indicate the direction of flux motion within a region of polarity outlined by contour at $\pm$100G . We normalized the magnitude of velocity vectors to 0.6 km/s such that the features which move at far less velocity will be clearly visible. From the Figure~\ref{Fig_vel}a, the velocity field in the negative polarity show a flow pattern towards east and that in positive polarity indicates a flow pattern towards west. These flow patterns around the PIL of region manifests shearing motion. This motion is known to effectively transform the initial potential configuration to sheared arcade, which then undergo reconnection at the converging region to form sigmoidal configurations \citep{Amari2003}. In the subsequent instances, the velocity pattern in both the polarities indicate flows are converging towards each other (within the orange oval). These motions about the PIL causes the flux cancellation process. \citet{Chae2004} reported that submerging opposite polarities are the sites of cancelling magnetic features. Besides the sigmoid formation by the slow shear and converging motions, a filament channel is observed in AIA 304~\AA~images from January 5, which implies a flux rope topology where dipped field lines support the filament material.

\subsection{Filament initiation and onset of main phase reconnection}
\label{subsec_filament}
\begin{figure*}[!ht]
 	\centering
 	\includegraphics[width=.99\textwidth,clip=]{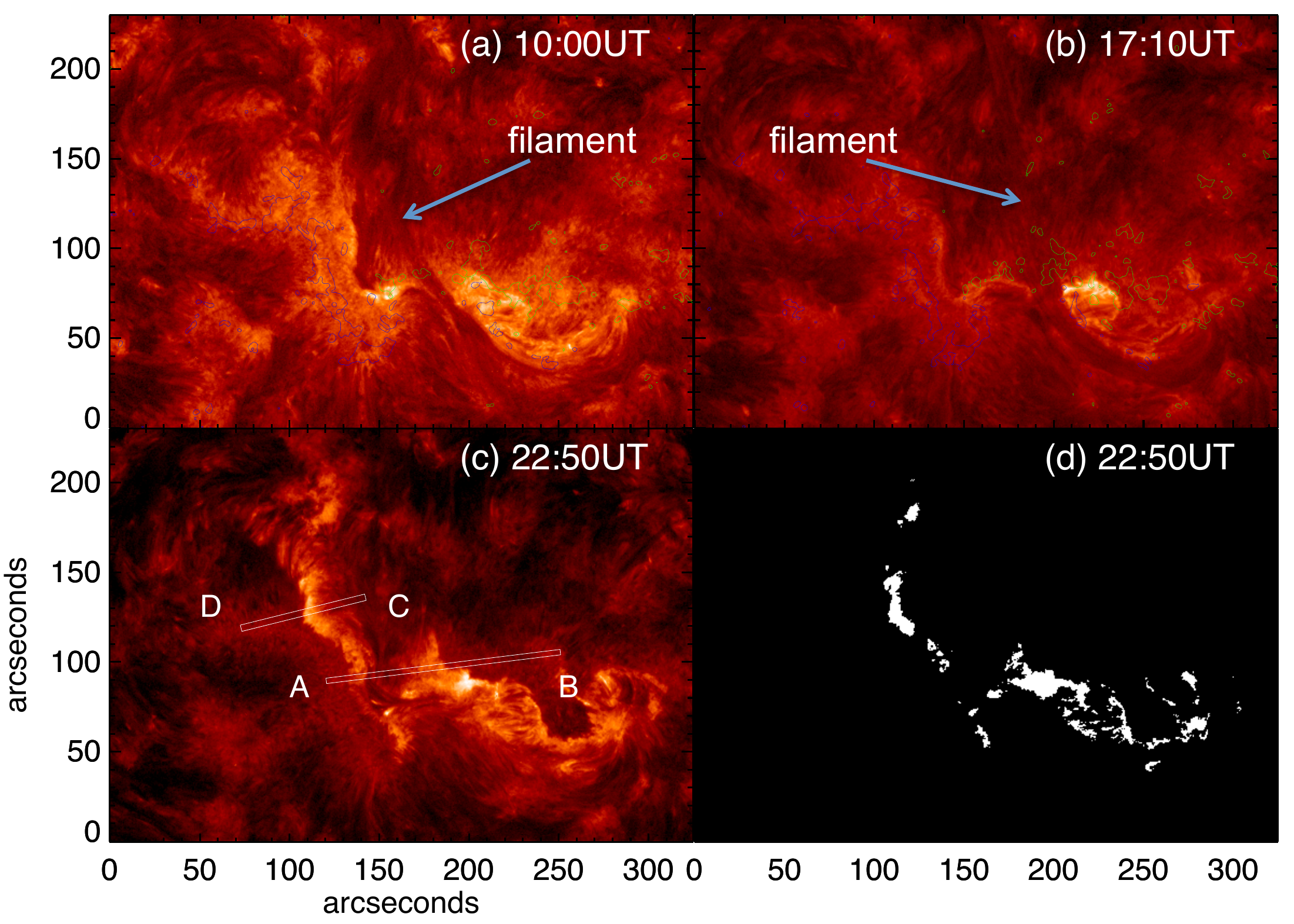}
 	\caption{AIA 304~\AA~ observations on 6 January 2014 showing filament rise motion during flux cancellation phase. (a) At 10:00UT the filament was seen as an integral part of the sigmoidal structure. (b )At 17:10UT the filament appeared rising upwards. Contours of line-of-sight magnetic field observations at $\pm$100G are overplotted in both panels (a \& b).(c) At 22:50 UT the flare ribbons were observed after the filament eruption. Slices AB and CD were placed to track the ribbon motions. (d) Flare ribbon mask generated from panel (c) is shown and be used to compute the mean magnetic field strength swept by flare ribbons at 22:50UT.} 
 	\label{Fig_filament_rib_mask}
 \end{figure*}

 In Figure~\ref{Fig_filament_rib_mask}a, the filament is seen to be supported by the sigmoidal structure. This filament channel exhibits some dynamical activity early on January 6, co-temporal with the net flux decrease by flux cancellation process. During this period, the filament channel appears to rise higher in height, especially the north lobe segment. These observations imply that continued shearing and converging motion of the fluxes lead to the onset of filament rise motion by the tether-cutting reconnection from around 05:35 UT.  In figure~(\ref{Fig_filament_rib_mask}b), the rising filament is captured in AIA 304~\AA~ image at 17:10UT. After 21:00UT the filament becomes invisible to AIA 304~\AA~ passband as it had risen to higher altitude and eventually erupts at 21:32 UT.
 
\begin{figure*}[!ht]
	\centering
	\includegraphics[width=.99\textwidth,clip=]{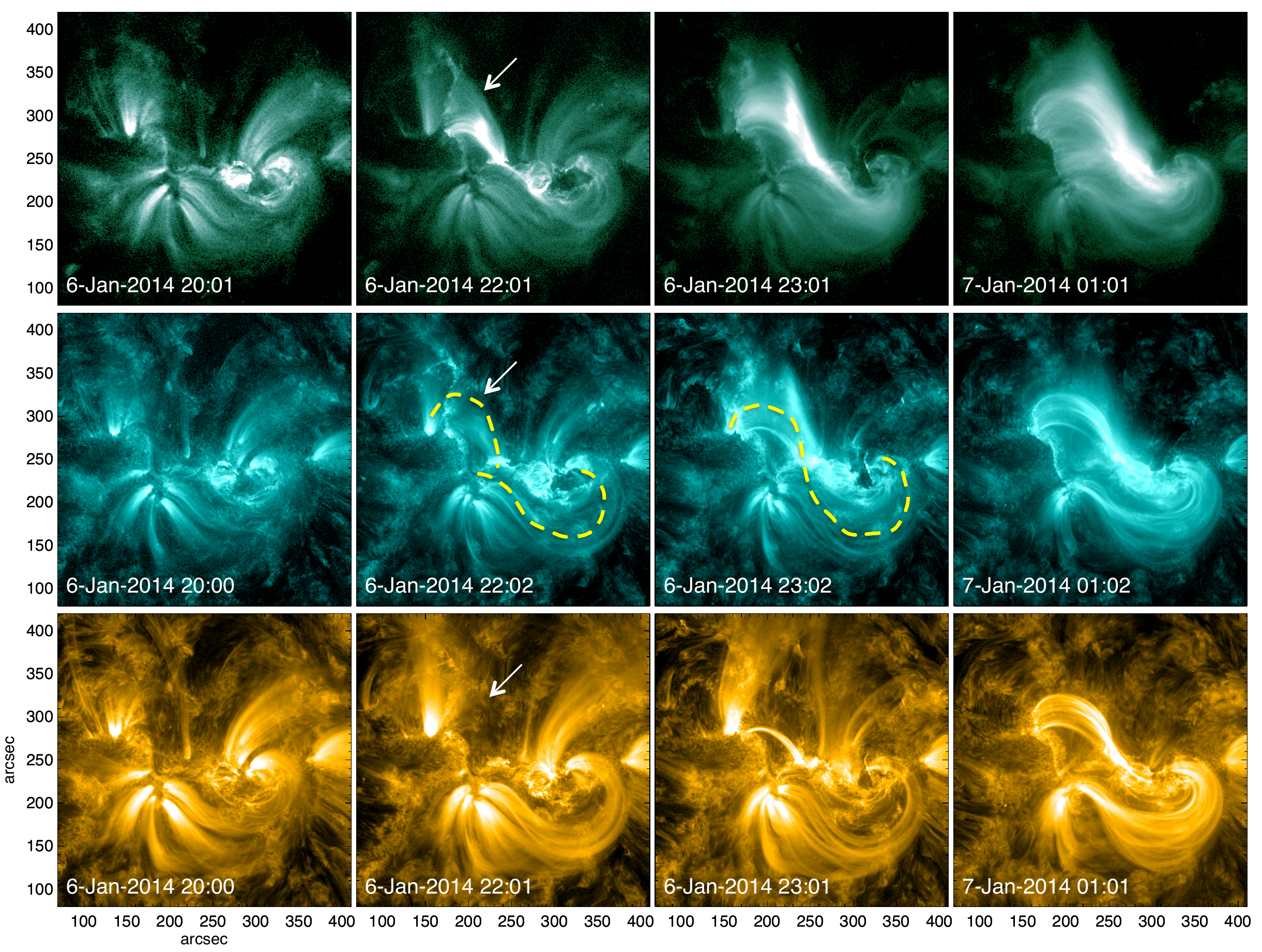}
	\caption{Observations showing onset of eruption by reconnection of inverse-J shaped loops. First row: Images in AIA 94\AA~waveband, Second row: Images in AIA 131\AA~ waveband, Third row: AIA 171\AA~waveband. During reconnection, hot loop system develops and is shown by white arrows in second column images.Yellow dashed lines traces the inverse-J shaped loops and S-shaped loop in the second and third AIA 131\AA~ images (second row) respectively.} 
	\label{Fig_erup_mos}
\end{figure*}

The morphological transformation in the sheared arcade during the filament eruption(i.e.,from 20:00UT to 01:00UT on January 6) is demonstrated with AIA EUV observations in Figure~\ref{Fig_erup_mos}. The first and second row images are from two hotter passbands AIA 94~\AA (~6.4MK) and AIA 131~\AA (~10MK) respectively, whereas third row images are from relatively cooler passband AIA 171~\AA (~0.6MK). The bundle of two sheared inverse J-shaped loops are evident in the second column images of Figure~\ref{Fig_erup_mos} and also traced in AIA 131\AA~ image (middle row) as yellow dashed lines. These two loops eventually get reconnected to form the long Inverse S shaped loop as depicted in the subsequent 3rd and 4th column images of Figure~\ref{Fig_erup_mos}.  Observations of continuous flux cancellation by converging and shearing motions suggests the tether-cutting reconnection model \citep{Moore1980}, where the inverse J-shaped field lines come closer and reconnect to form long inverse S-shaped loops. Due to this tether-cutting reconnection, we expect that the sigmoidal field lines will be heated to higher temperatures ($>$2-20MK) and the resultant enhanced emission is captured in hotter AIA 94\AA , AIA 131\AA~ and AIA 193\AA~ pass bands, whereas  the cooler AIA 171\AA~ pass-band couldn't detect the emission enhancement from the heated plasma from reconnection. This scenario is pointed by white arrows on second column images. From the AIA movies of 94\AA~ and 131\AA, we observed the initial enhanced brightness due to reconnection at 21:32 UT on January 6, which is presumed to be the time of onset of main phase reconnection between two inverse J-shaped loops. The addition of axial flux strengthen the flux-rope and is seen as expanding CME in the north-west side, which is not visible directly in EUV images but clearly in the running difference images of AIA 171\AA. Further, the repetitive reconnection continues in the sheared arcade and the emission of resultant heated plasma is suggested to be responsible for brightening of the observed transient sigmoidal structure. 

\subsubsection{Flare-ribbon velocity and reconnection rate}
\begin{figure*}[!ht]
 	\centering
 	\includegraphics[width=.92\textwidth,clip=]{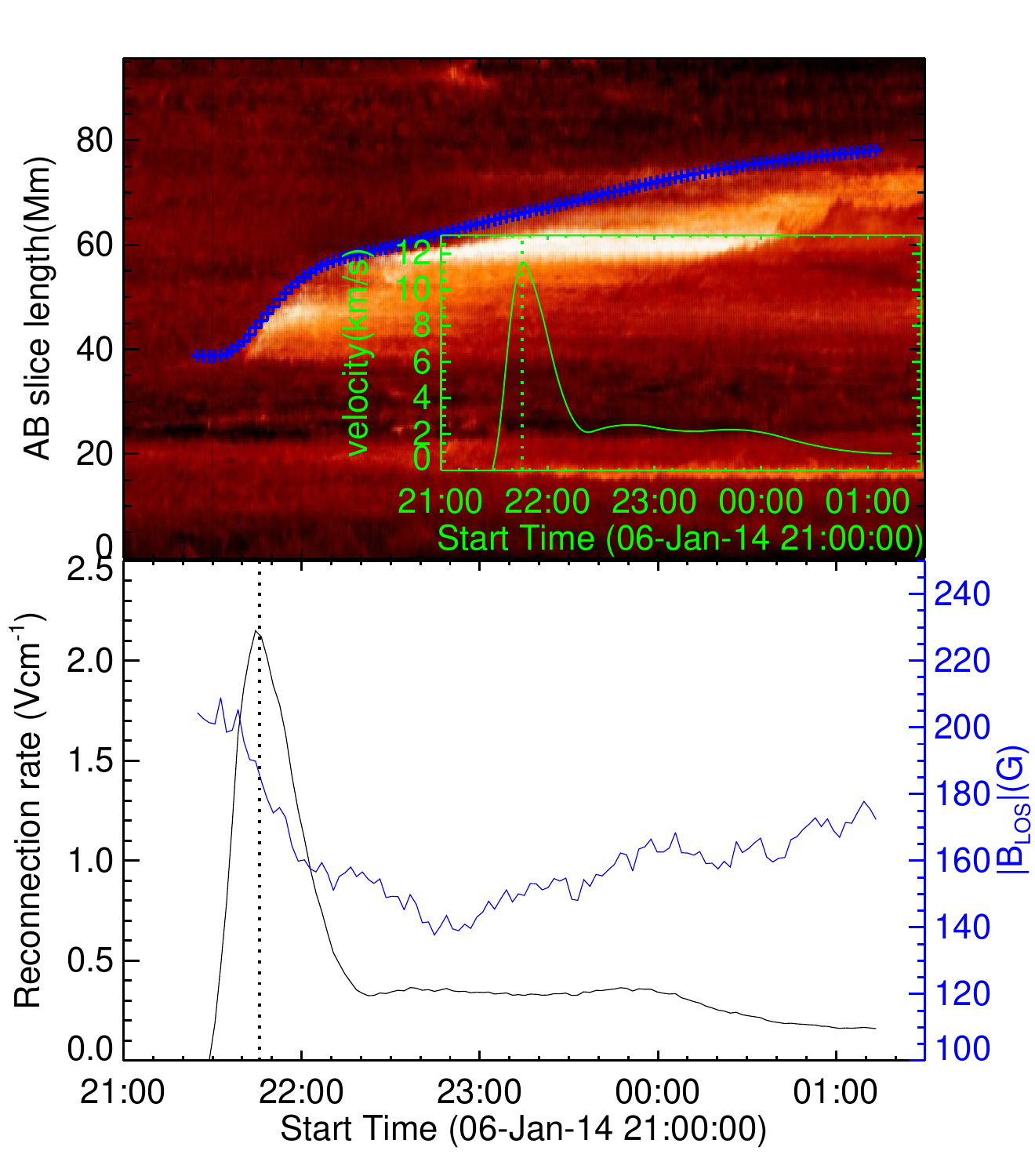}
 	\caption{ Top panel: space-time plot of slice AB (Figure~\ref{Fig_filament_rib_mask}c) generated from AIA 304\AA~ images. The bright section traces the motion of flare ribbon. Blue ``+" symbols represents the spline fit to the distance data points and inset(green) shows the temporal profile of ribbon velocity derived from time derivative of spline-smoothed distance profile. Bottom panel: Temporal evolution of local reconnection rate (black) and absolute mean LOS magnetic field strength(blue) swept by the flare-ribbons. Velocity and reconnection rate profiles are temporally well correlated and both peaks at 21:47 UT (vertical dotted line) corresponding to the flare peak time.} 
 	\label{Fig_ribbon_vel_rec_rate}
 \end{figure*}

We observed the flare-ribbons and their separation from 21:40 UT on January 6 to about 01:00UT on 7 January in AIA 304~\AA~ passband images. To track the flare-ribbon motion, we used the AIA 304~\AA~ images at the cadence of 2 minute and determined the ribbon separation velocities using slices at different directions along ribbon motion  as shown in figure~(\ref{Fig_filament_rib_mask}c). The stack-up of one such slices(AB) with time is shown in top panel of figure~\ref{Fig_ribbon_vel_rec_rate}. To smooth out fluctuations due to measurement uncertainties, spline smoothing was applied to the distance data points which are identified to be tracked front edge of the flare-ribbon. The spline fit to the flare-ribbon distance profile is marked with blue ``+" symbols. The required temporal evolution of the apparent ribbon velocity was determined as the time derivative of the spline-smoothed distance profiles and is shown as an inset(green) in top panel of Figure~\ref{Fig_ribbon_vel_rec_rate}. 

 \citet{Forbes1984} and \citet{Forbeslin2000} showed that the local reconnection rate i.e., the rate at which magnetic field lines are carried into the reconnection site, is given by the coronal electric field E at the reconnection site. They derived a simple relation between the local reconnection rate and the apparent flare-ribbon separation speed $v$. It is given by $ E=B_{n}v $, where $ B_{n} $ is vertical component of magnetic field underlying flare-ribbon location. To determine the photospheric magnetic field strength, first we generated the flare-ribbon mask by giving the intensity threshold ($ \geq 10^{2.5} DNs^{-1}pixel^{-1} $) to AIA 304~\AA~ image. The masks are generated at a cadence of 2 minute for every AIA 304~\AA~ image. One such example mask is shown in panel~\ref{Fig_filament_rib_mask}d. Later the masks are multiplied with the near simultaneous co-registered LOS magnetograms to get the underlying magnetic field swept by flare-ribbon front. To reduce the uncertainty, we excluded the magnetogram pixels having value less than 20 G in the computation of absolute mean of photospheric magnetic field. The temporal profiles of mean magnetic field strength and local reconnection rate are plotted in the bottom panel of Figure~\ref{Fig_ribbon_vel_rec_rate} . The velocity of flare-ribbon (one moving towards west) and local reconnection rate reach their maximum values of 11.52 kms$^{-1} $ and 2.14 Vcm$^{-1} $ respectively at 21:47 UT($\pm1$minute) and is co-temporal with flare peak time. Our results are comparable with the maximum reconnection rates obtained for C-class eruptive flares in \citet{Hinterreiter2018} and lesser than the maximum reconnection rates obtained for strong eruptive flares \citep{Miklenic2007,Hinterreiter2018}. 
 
\subsection{Post eruption evolution and the CME}
\subsubsection{Light Curves}
\label{subsec_lc}
\begin{figure*}[!ht]
	\centering
	\includegraphics[width=.99\textwidth,clip=]{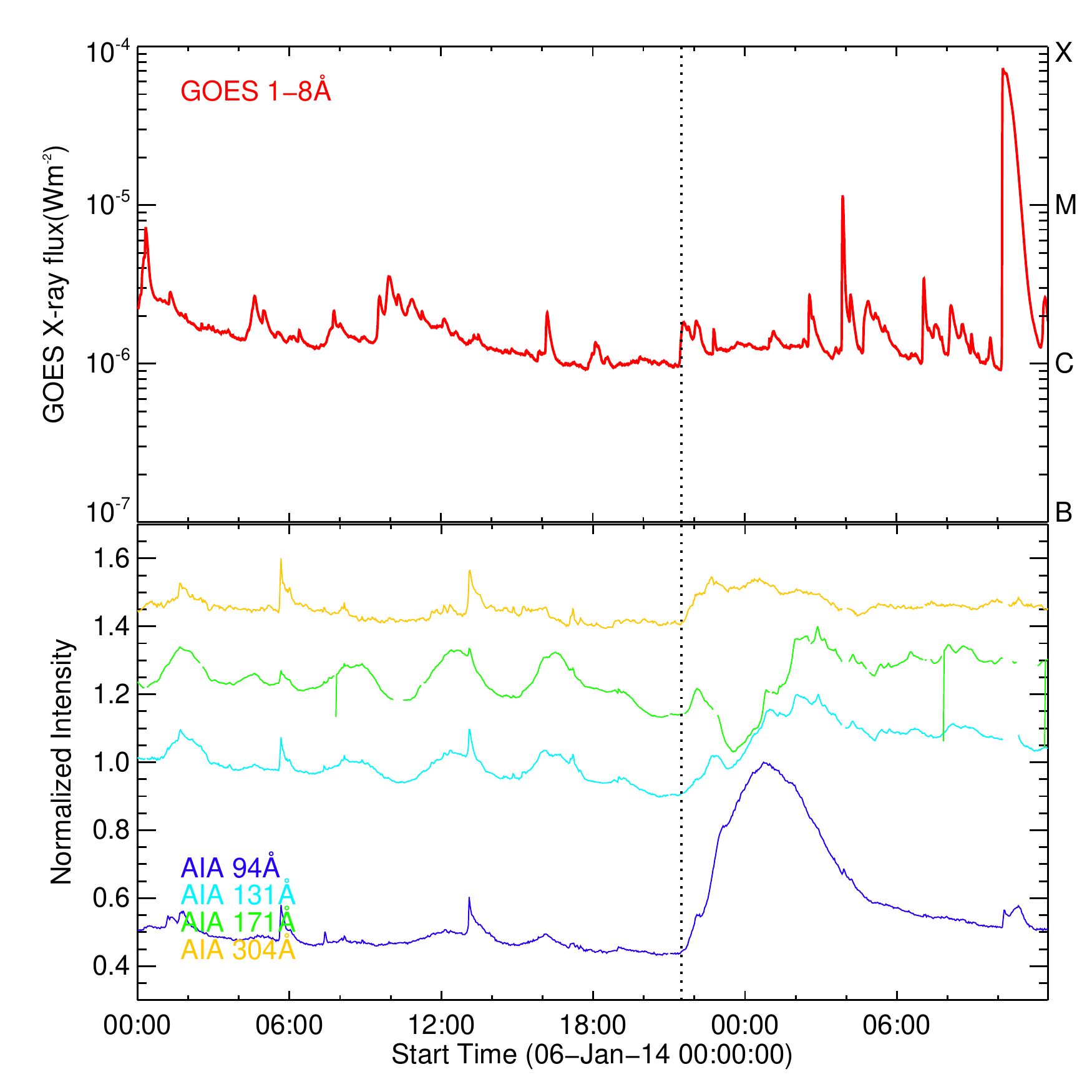}
	\caption{Top panel:Temporal evolution of disk integrated GOES SXR flux in 1-8\AA. Bottom panel: Light curves of AR 11942 in EUV wavebands of AIA. Vertical dotted line marks the onset time (21:32UT) of main phase reconnection. Vertical shifts (relative) to each light curve were given for better presentation. Note that the AIA 304, 94 light-curves start increasing with the onset of reconnection at 21:32UT on January 6 and peaks at 22:40UT on January 6, 00:44UT on January 7 respectively.} 
	\label{Fig_lc}
\end{figure*}

Evolution of GOES soft X-ray (SXR) flux in 1-8\AA~ and the light curves of AR 11942 obtained from AIA  images of different wavelengths are shown in the top and bottom panels of Figure~\ref{Fig_lc} respectively. The GOES SXR background is already above C-class due to the X-ray flux from other ARs and there was a sudden small flux enhancement corresponding to the EUV flux from AR 11942 at 21:32 UT(black vertical dotted line). Unlike many eruptive flares, this SXR enhancement as C1.0 flare remains an hour and is regarded as weak flare compared to the long EUV activity.  A time delay between peak emissions of different EUV wavelengths were noticeable from bottom panel of Figure~\ref{Fig_lc}. All these light curves were normalized to their maximum intensity and were vertically shifted by some amount to avoid overlapping of these curves.

The time profile of  AIA 94\AA~ starts rising from 21:32UT and reaches gradually to peak emission at 00:44UT on January 7 and then follows a slow decay phase. The emission from hot coronal inverse-S shaped loops were clearly captured in  AIA 94\AA~and AIA 131\AA. 131 \AA~filter covers Fe VIII ($LogT \approx 5.6$), Fe XX ($LogT\approx 7.1$), and Fe XXIII ($LogT\approx7.2$) which are sensitive to low temperature at 0.6 MK and high temperature at 10 MK as well. The 131\AA~ has peak emission from the AR at 2:54 UT on January 7 and is delayed by 130 minutes compared to AIA 94\AA. The additional heating provided by the slow magnetic reconnection in the gradual phase creates the delayed peak emission in AIA131 \citep{Liu2015}. Light curve of AIA 304 shows an impulsive rise and reaches quickly to its peak at January 6, 22:40 UT even before AIA 94 \textbf{light curve}. This shows that chromosphere and transition region immediately responded to heating to at least 1MK \citep{Chamberlin2012} and its wider peak is accounted for by different cooling timescales of the heated coronal loops. A dip in the AIA 171 profile during main phase reconnection is mainly due to the hot reconnection loops which are opaque to the cooler passband of AIA 171\citep{Dai2018}. Later these  loops slowly started to appear in AIA 171 and warm coronal emission peaks at 2:52UT on January 7, which is mainly due to the long cooling process of the late-phase loops as proposed in \citet{Liu2013}. It is worth to note that the AIA 171 and AIA 131 peak emissions (both AIA131 and AIA171 passbands are sensitive to cooler temperature of $\approx0.6MK$) got delayed by  about 130 minutes compared to AIA 94\AA~.  Also, such time delays in peak emissions of light curves of strong eruptions (occurred from ARs for example \citet{Vemareddy2014}) are quite smaller ($\approx$ 10min). Though studies like \citep{Liu2015,Dai2018} also observed the large time delays between peak emissions in EUV light curves of AR during confined flare events, we need more  comparative studies between strong  and weak eruptive ARs to verify our result.

\subsubsection{CME detection}

\begin{figure*}[!ht]
	\centering
	\includegraphics[width=.99\textwidth,clip=]{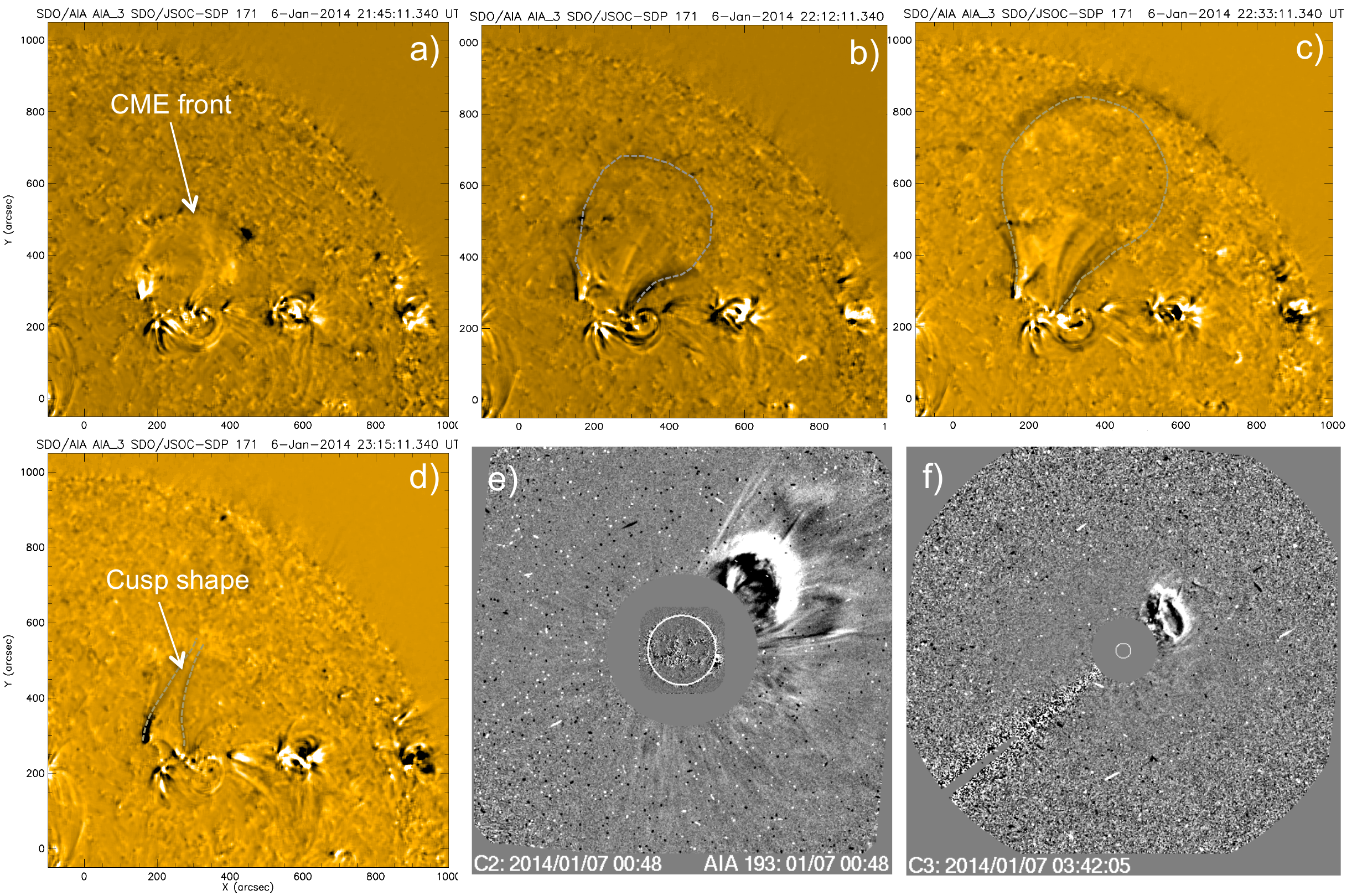}
	\caption{Image sequence of CME flurope eruption from AR 11942. (a-d): Time difference images of AIA 171\AA~observations. To enhance the CME front, the actual observations are smoothed before taking time difference. Cusp-shape loop structure is visible below leaving CME blob and is shown in panel(d). (e-f): Further propagation of the CME in LASCO C2, C3.  (See attached animations for this figure) } 
	\label{Fig_cme}
\end{figure*}

The AR produces an eruption after the formation of sigmoid. As the eruption occurred on-disk, the projection effect reduces the contrast and consequently hampers the faint eruption on-disk to be seen clearly in  EUV/AIA intensity images. But on applying running difference technique to AIA 171\AA~images, we can observe the propagation of the CME flux rope unambiguously. Figure~\ref{Fig_cme}(a-d) represents the sequential expansion and eruption of CME from the source AR in north-west direction. To filter out high frequency features and to improve contrast, we apply high degree of smoothing to the 5 minute running difference images of AIA 171\AA.  The traced grey line in the panel~\ref{Fig_cme}c represents the stretching and expanding CME front, while still having connections to its source AR. 

 Panel~\ref{Fig_cme}d represents the post-eruption scenario, as the CME  moves away from the limb the stretched loops shrink to form the cusp shaped structure below. This eruption scenario matches with the catastrophic CME model. This leads to the stretching up of overlying loops and as a result opposite field loop lines started getting close to each other to a cusp-shape structure. As the loops came in contact with each other, the current sheet forms in between and the subsequent reconnection in the current sheet leads to form post-CME loops below. Panels~\ref{Fig_cme}(e-f) display the observed CME in LASCO C2 and C3 FOV respectively. 

\subsubsection{CME kinematics}

\begin{figure*}[!ht]
	\centering
	\includegraphics[width=.92\textwidth,clip=]{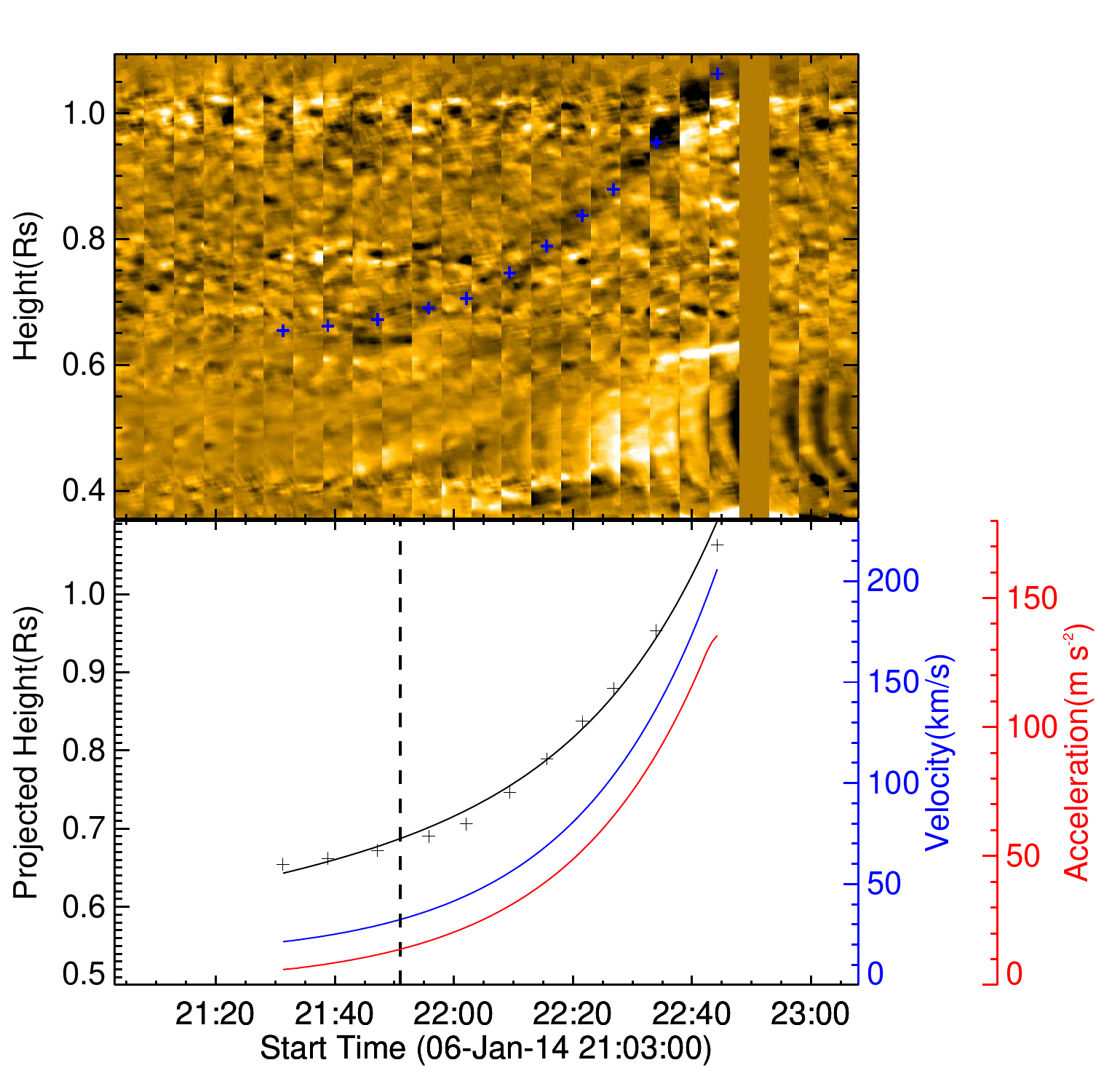}
	\caption{ Top panel: space-time stack plot of the slit placed across the expanding CME front observed in AIA 171\AA~ difference images. Bottom panel: Projected height-time plot of the expanding CME front. ``+" symbols (in both panels) traces the ascending CME front. Black solid curve is model fit to the height-time data, and blue (red) curve is the derived velocity (acceleration) of the CME front.Vertical dashed line (21:51UT) represents the onset of rapid-acceleration phase and it divides the eruption occurred into two phases, the slow-rise phase of duration 21:32 - 21:51UT with an average velocity of ~26 km/s followed by rapid-acceleration phase from $ 14ms^{-2} $ to $ 134ms^{-2} $ during 54 minute duration.} 
	\label{Fig_cme_ht_vel}
\end{figure*}

We performed on-disk kinematic analysis of the CME using 5 minute running difference images of AIA 171\AA. On careful inspection of running difference images, we identified the leading edge of CME fluxrope and placed a slit along its ascending direction. Since the eruption occurred on-disk, the height refers to the projected distance on the disk from the center of the sun. Thus the resulting on-disk velocity  of CME fluxrope is also approximated. The space-time image is shown in top panel of Figure~\ref{Fig_cme_ht_vel} and blue ``+" symbols in the image represents ascending CME front. 
 We used a model containing the linear term to treat the slow-rise phase and exponential term to account the rapid-acceleration phase as described in \citet{cheng2013} and is given by 
$h(t) = C_{0} e^{(t-t_{0} )/\tau} + C_{1} (t-t_{0} ) + C_{2}$,  where $h(t)$ is height at a given time t, and $\tau$, $t_{0}$, $C_{0}$, $C_{1}$, and $C_{2}$ are free coefficients. This model has two distinct advantages, firstly a single function describes the two phases of eruption effectively and secondly it provides a convenient method to determine the time of onset of rapid-acceleration phase ($T_{c}$). The onset of rapid-acceleration phase is defined as the time at which the exponential component of velocity equals to its  linear component as $T_{c} = \tau ln(C_{1}\tau /C_{0} ) + t_{0}$. 
We used \texttt{mpfit.pro} to fit the height-time data.  From the fit, $T_{c}$ is determined to be 21:51UT (indicated by vertical dashed line  in bottom panel of  Figure~\ref{Fig_cme_ht_vel}). 

Further, the numerical derivative method is applied to calculate the velocity and acceleration profiles (e.g., \citet{Zhang2001,Zhang2004}). The derived velocity (blue) and acceleration (red) profiles are over-plotted in bottom panel of Figure~\ref{Fig_cme_ht_vel}. The duration of slow-rise phase is between 21:32 - 21:51UT with an average velocity of ~26 km/s. By 21:51 UT, the CME front acquired fast-rise motion and velocity(acceleration) increased from 36 km/s($14ms^{-2}$) to 206km/s($134ms^{-2}$) in 54 minutes, with an average acceleration of $\approx 55ms^{-2}$. These are quite different from strong eruption cases, for example \citet{Zhang2012} showed the CME flux rope in slow-rise phase has the average velocity of about 60 km/s and in fast-rise phase, it reached the terminal velocity of about 700km/s in 6 minutes with an average acceleration of $1600ms^{-2}$. Similarly in another example of quite stronger eruption event presented in \citet{Cheng2014}, the CME flux rope had average velocity of about 35km/s during its slow-rise phase and in rapid-acceleration phase it reached the terminal velocity of about 300km/s in 23 minutes with an average acceleration of $ 200ms^{-2} $. Thus the eruption under study is quite a slower one and we opine that, the slow-rise phase of CME flux-rope is due to the tether-cutting reconnection and the fast-rise phase is accounted for by the catastrophic behaviour of the flux rope system and/or torus instability \citep{Demoulin2010,ChenH2018}. It will be discussed further in the section~\ref{summ}. 

The sigmoid eruption is observed as CME visible in SOHO/LASCO which the CME catalog listed as two CMEs appearing in LASCO C2 field-of-view (FOV) initially at 23:12UT and 23:24UT at position angles 275 and 300 degrees respectively. We believe that these could be different parts of the single CME structure, named as part1 and part2. The part1 was identified to be expanding CME front as observed in AIA 171\AA~ difference images of Figure~\ref{Fig_cme}(a-d)(also refer animation) and further it is tracked by coronagraph C3 until January 7 at 00:42 UT up to a height of 9 R$_\odot$. The part2 was tracked by coronagraph C3 until January 7 at 8:00 UT upto  a height of 22 R$_\odot$. The height$-$time plot available at the SOHO/LASCO CME catalog shows the linear speeds of  part1 and part2 are ~722 km/s and ~442 km/s with their angular widths 15 degree and 100 degrees respectively. \citet{Gopalswamy2001} found that the average speed of CMEs associated with decameter-hectometric radio type II bursts to be 960 km/s  and the average width to be 66 degree. In other words, the strong geo-effective CMEs are those having speed and width greater than 960 km/s and $66^\circ$ respectively. Therefore, the CME event under study is considered as slow/weak CME. 

\subsubsection{Relaxation of sheared core field}
Past Observations \citep{Sterling2000, Pevtsov2002,Liu2010} have  shown that eruptive ARs, which initially display a sigmoidal structure, evolve into a post-eruption phase that consists of field lines with a cusp-like shape. The newly formed field lines of sigmoidal structure are highly sheared, containing much non-potential magnetic energy. The reconnection occurs in sheared core field \citep{Moore2001} helps in the release of such contained free magnetic energy. The dissipation of free magnetic energy causes the sheared field to relax, giving rise to the contraction of  loops\citep{Ji2007}. Such a process of relaxation of sheared field lines is represented by typical AIA 94~\AA~passband images.  By 01:00 UT, the CME had moved away from the sheared core field, the enhanced brightness in the core field indicates the continuous reconnections allowing the long sheared field lines to relax as seen in EUV observations (figure not shown).

\subsection{Thermal Evolution}
Emission measure and thermal structure of sigmoid is explored by applying Differential Emission Measure (DEM) analysis to six EUV passbands of SDO/AIA. We used \texttt{xrt\_dem\_iterative2.pro} in Solar Software with modifications to work with AIA data. The code was initially designed for Hinode/X-ray Telescope data \citep{Golub2004, Weber2004}. However, \citet{Cheng2012} tested this code extensively on AIA data by  studying the different thermal properties of multi-structure CME. The DEM maps of sigmoid are constructed and the emission measure (EM) and DEM weighted average temperature ($\bar{T}$) are derived using the following definitions, 
\begin{equation}
\bar{T} = \int DEM(T) T dt/ \int DEM(T) dt; \,\,\, EM = \int DEM(T) dt
\end{equation}
\begin{figure*}[!ht]
	\centering
	\includegraphics[width=.99\textwidth,clip=]{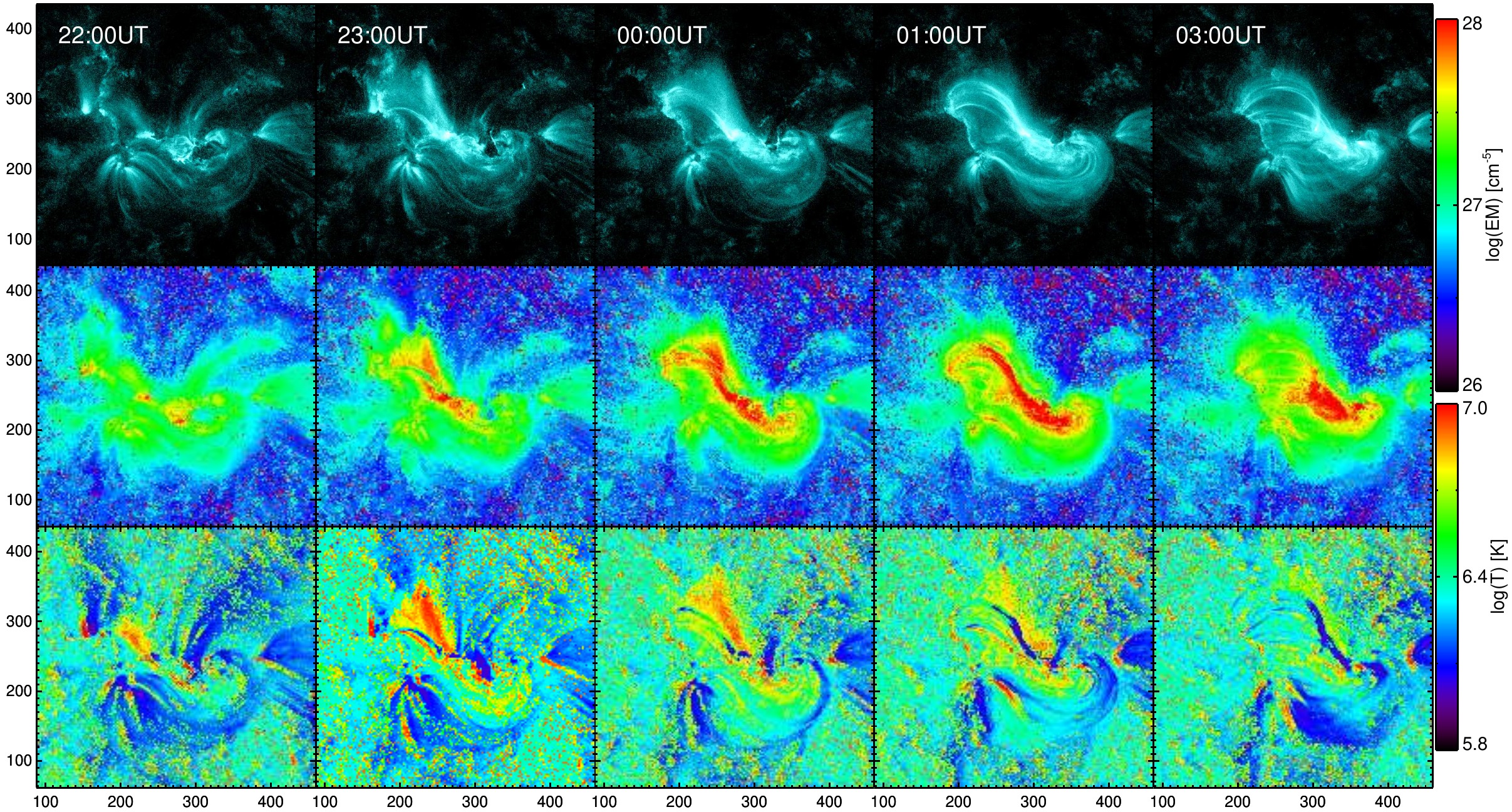}
	\caption{Top row: AIA 131~\AA~waveband images represent the typical evolution of sigmoid. Middle row: Maps of EM distribution corresponding to first row panel. Bottom row: maps of T distribution.  In the third and fourth columns, the EM and $\bar{T}$ distribution reaches maximum in and/or around bright reconnected loops. } 
	\label{Fig_em_tm}
\end{figure*}

where integration is performed in  temperature limits of $5.7<LogT<7.1$. In Figure~\ref{Fig_em_tm}, we plot the maps of $EM$ and $\bar{T}$ representing the distribution of total emission measure and average temperature of sigmoidal structure in middle and bottom panels respectively, while the corresponding AIA 131\AA~images are displayed in top panel. The distribution of EM and $\bar{T}$ in sigmoidal structure is higher than that of the quiet region surrounding it in all panels. As the magnetic reconnection had already started at 21:32UT, the enhanced brightness is clearly seen in AIA 131\AA~image \textbf{taken at} 22:00UT at the reconnection site. Similarly in the  same time slab, the enhanced EM and $\bar{T}$ distribution is seen in the reconnection region. The released energy due to magnetic reconnection heats up the plasma resulting in the emission. At 23:00UT, the hot post-eruption loops forms below resulting in increase of temperature and EM in and around those loop lines. At 00:15UT on January 7, the EM and $\bar{T}$ distribution reaches maximum in and/or around bright reconnected loops. The $EM$ reaches the maximum value of $10^{29} cm^{-5}$ and average temperature ($\bar{T}$ ) reaches upto $10^{7.09}$ K. However, the maps are scaled optimally for better contrast. After 1:00UT, the decrease in the EM and $\bar{T}$ distribution starts as evident in time slab of 3:00 UT. 

\begin{figure*}[!ht]
	\centering
	\includegraphics[width=.99\textwidth,clip=]{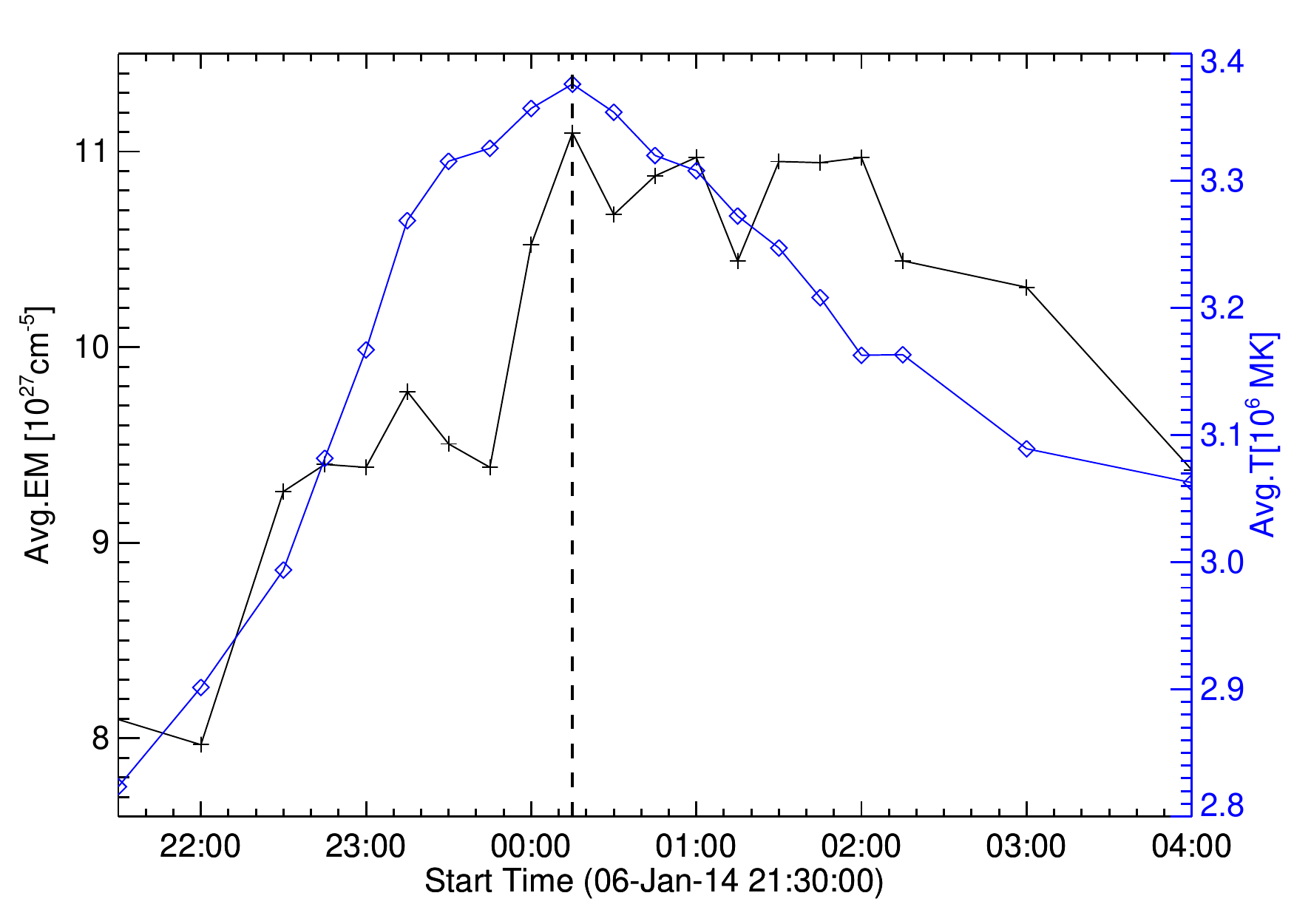}
	\caption{ Temporal evolution of average $EM$ and $\bar{T}$ in the AR. The dashed vertical line represents the peak time (00:15UT) of both curves.The $\bar{T}$ curve gradually declines from its peak but $EM$ profile shows steady emission till 2:00UT  representing strong emissions from cooling loops and then declines slowly. } 
	\label{Fig_em_tm_curves}
\end{figure*}

The DEM reconstruction is an ill-posed problem and several errors arise from uncertainties in response function, background determination, radiative transfer effects etc., \citep{Judge2010}. Hence we concentrated on the temporal variation of EM and $\bar{T}$ rather than their exact values. We estimated the average values of EM and $\bar{T}$ of the sigmoidal structure with the same field-of-view as shown in the Figure~\ref{Fig_em_tm} during the time-period from 21:30UT to 04:00UT at an interval of 15 minute. The temporal evolution of these curves are shown in Figure~\ref{Fig_em_tm_curves}. As the CME moves away from the solar disk (by 22:50UT), the hot post-eruption loops form underneath. Due to this, the temperature and EM curves start to rise further after the eruption. The EM curve  in black and $\bar{T}$ curve in blue, both follows the same trend  reaching their peaks at 00:15UT as indicated by the vertical dashed line in Figure~\ref{Fig_em_tm_curves}. After reaching peak, the average $\bar{T}$ curve shows the gradual declining phase, whereas the EM curve shows the steady phase till 02:00UT and then declines. This shows that the emission from cooling loops are quite strong enough and  as comparable to that of hotter loop structures during that time period. Also, these curves follows the light curves of sigmoid as described in previous section~\ref{subsec_lc}.

\begin{figure*}[!ht]
	\centering
	\includegraphics[width=.99\textwidth,clip=]{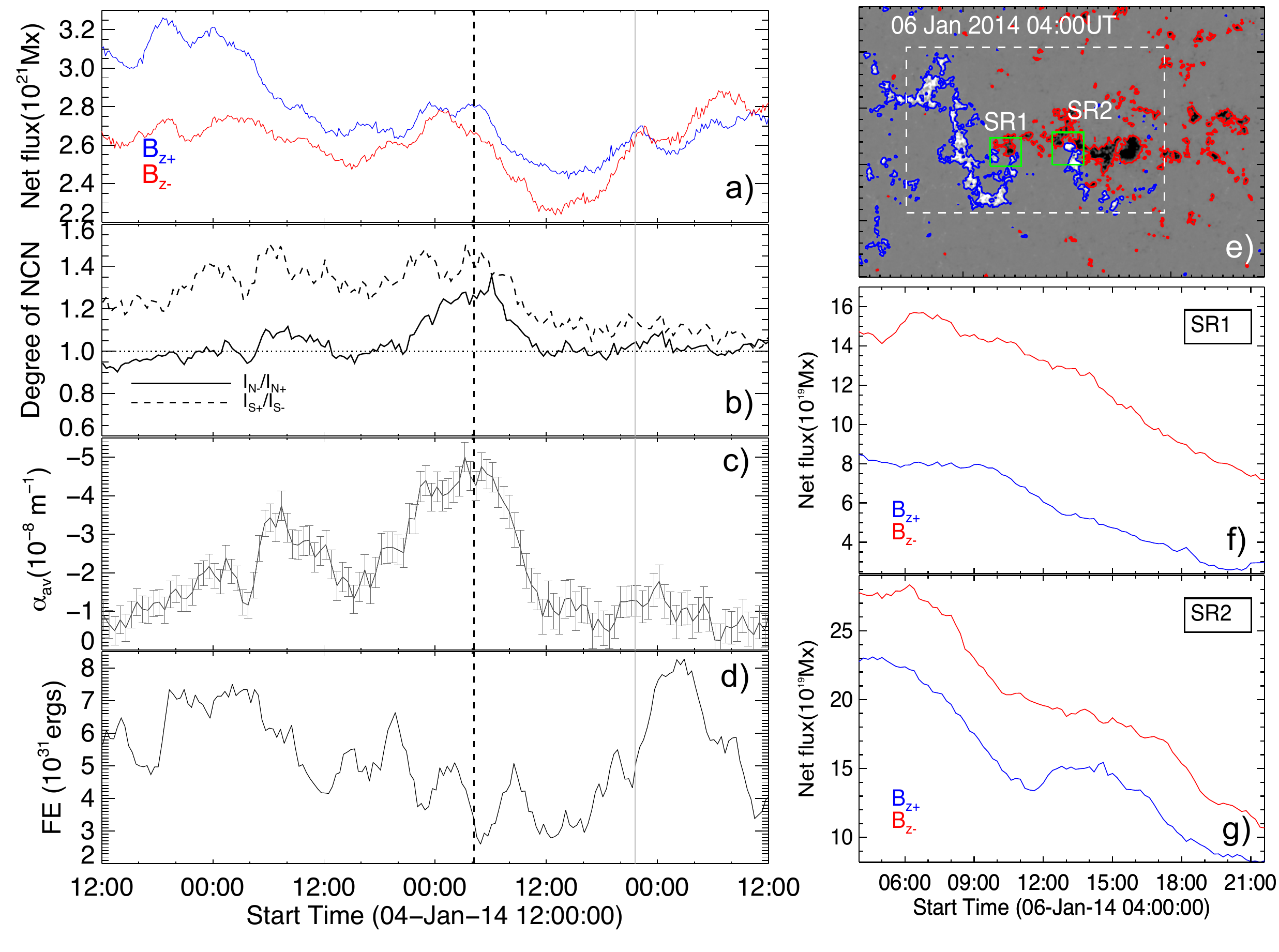}
	\caption{Time evolution of magnetic parameters in the region enclosed by white dashed rectangle in panel (e). a) net flux, b) degree of net current neutralization, c) $\alpha_{av}$ d) free magnetic energy at the photosphere. Vertical dashed line (04:00UT) marks the time when net flux starts decreasing till the time of eruption (21:32UT) indicated by vertical grey line. Also, note the currents are non-neutralized before and during the time of eruption. e-g) Net flux evolution in converging sub-regions SR1, SR2. Note that the net flux in the sub-regions show decreasing flux content over time.  } 
	\label{Fig_mag_time_evol}
\end{figure*}

\subsection{Evolution of photospheric magnetic parameters}
 We studied the temporal evolution of different magnetic measures for 3 days ( 2.4 days prior to eruption) using the HMI vector magnetograms(SHARP series) at the cadence of 12 minute. Generally the magnetogram measures  are area-dependent, hence these parameters are computed by choosing an area within the active region(AR) including eruption region with minimum flux-imbalance(i.e., net flux/total unsigned flux $ < $ 3\%). Moreover, a careful manual inspection of field lines connectivity in  AIA 171~\AA~ images results in the exclusion of certain regions of  AR  connecting to neighbouring  ARs and thereby restricting the area only to the eruption region with minimum field lines connecting outside of it. The area enclosed by the white dashed rectangle in Figure~\ref{Fig_mag_time_evol}e is used to compute the following magnetogram measures.

The total unsigned flux of AR is defined as $\Phi =\sum |B_z| dA$ , where $dA$ is the area of an observation pixel. The flux is computed from the pixels having magnetic field strength higher than 50 G. Since the measurement error is within 15G, we have chosen the threshold three times larger than that and hence the error in our estimation of flux is negligible. The temporal evolution of net flux is shown in Figure~\ref{Fig_mag_time_evol}a. The frequent emergence and cancellation of fluxes are evident in AR during its disk passage from January 4 to middle of January 7. Notable observations are from 4:00UT on January 6 (as indicated by vertical dashed line in Figure~\ref{Fig_mag_time_evol}), where the net flux starts decreasing  corresponding to flux cancellation and such decreasing continues till the onset (21:32UT; indicated by vertical grey line in Figure~\ref{Fig_mag_time_evol})  of the main phase reconnection. During this time period, the filament exhibits dynamical evolution and rise motion initiated from 05:35 UT. The regions of flux cancellation during this time period are studied in section~\ref{subsec_shear_fc}. Careful scrutiny of simultaneous observations of magnetograms and AIA images (observed several episodes of EUV brightenings and also converging motions in small regions) led to select two sub-regions namely SR1 and SR2 as indicated in Figure~\ref{Fig_mag_time_evol}(e-g). The net flux in these regions is plotted in the bottom panels, which show continuous decreasing evolution. We thus suggest from these observational evidences that the persistent slow shearing and converging motions about the PIL played prime role in the cancellation of fluxes leading to the initiation and eruption of the sigmoid configuration  \citep{Green2009,Green2011,Savcheva2012,Vemareddy2015,Vemareddy2017}.

 The vertical current density is computed  as  $J_z = 1/\mu_0  (\partial B_y/\partial x -  \partial B_x/\partial y)$, where $B_x$ and $B_y$ represents the horizontal component of magnetic field. For a given polarity, the current distribution contains both positive and negative values. We examined the degree of net current neutralization (NCN) in each polarity by obtaining the ratio of Direct Current (DC) and Return Current (RC) \citep{Torok2014}. The dominance of the signed currents decides the chirality of the magnetic field and so DC is also considered as the dominant current and RC as the non-dominant current. The DC and  RC  are computed for each polarity by integrating current density values of different signs separately. This has been done only for those pixels whose $B_x$, $B_y$ values are larger than 150 G and $B_z$ values are larger than 50 G to minimize the error in computation of currents. The ratio of  DC and RC  indicates the extent of departure from net current neutralization in any polarity. While selecting the integration area, we have not excluded the non-eruptive flux completely in AR so the values of $|DC/RC|$ of any polarity at any time interval are expected to be quite smaller than the actual values which are estimated by restricting the integration to the foot point area of eruptions \citep{Liu2017, Vemareddy2017a}. We found DC is positive in south polarity and negative in north polarity. The temporal evolution of  $|DC/RC|$ values in both polarity regions are plotted in  Figure~\ref{Fig_mag_time_evol}b. The $|DC/RC|$ values in south polarity region maintained well above unity. Although the evolution in north polarity has similar trend, the $|DC/RC|$ values fluctuates for the fact that the horizontal field in this polarity are not strong and the numerical differentiation may have artifacts rather to represent the reliable current distribution.  The non-neutralization current implies the eruptive behavior of AR in terms of flux rope models.  

Average alpha ($\alpha_{av}$), the proxy represents the twist of magnetic field lines in an AR, is computed using the equation given by $\alpha_{av}=  \sum [J_z (x, y) B_z(x,y) / |B_z(x,y)|]$ \citep{Pevtsov1994,Hagino2004}. The errors are estimated from the least-squared regression plot of $B_z$ and $J_z$ \citep{Vemareddy2012}.  The temporal evolution of  $\alpha_{av}$ along with its error bars are shown in Figure~\ref{Fig_mag_time_evol}c. The twist of field lines starts increasing from January 4 till early hours of January 6, consistent with coronal observations of the sigmoidal structure formation over a time scale of days as described in Figure~\ref{Fig_overall}. As soon as the flux cancellation starts i.e., at 4:00 UT on January 6, $\alpha_{av}$ decreases as the signature of disappearing flux from cancellation.

Further, the total magnetic energy and potential energy  in the coronal field is estimated using a Virial theorem equation \citep{Chandrasekhar1961,Molodensky1974,Low1982} as $E =\frac{1}{4\pi}\int(x B_x + y B_y) B_z dxdy$. Since the photospheric field is not force-free, the energy estimate by using this equation serves as proxy for energy content in the AR. Then proxy for magnetic free energy is estimated by taking the difference of total magnetic energy and potential energy. We used the whole AR for potential-field extrapolation but energy estimations were done for the region within the white dashed rectangle as shown in Figure~\ref{Fig_mag_time_evol}e.  As seen in panel~\ref{Fig_mag_time_evol}d, the free energy decreases corresponding to the increasing $\alpha_{av}$ till January 6. This may be due to emerging flux which increases the potential energy dominantly than the non-potential amount in the total energy. During the magnetic flux cancellation period (4:00UT - 21:32UT, 6 Jan), the free energy remains almost same with some undulations. It is important to note that the free energy during small C-class flare event is small and the field variations at different sub-regions camouflage the field contributing to the reconnecting flux. In such cases the expected variation of the free-energy is difficult to realize besides the intrinsic problems with the noisy transverse field observations. In conclusion, this study suggests that the energy release proceeded in a much long time duration, manifesting the onset of filament rise and eventual eruption driven by converging and canceling flux in the photosphere. 

\section{Summary and Discussion}  
\label{summ}
 In this paper, we investigated the sigmoidal eruption from AR 11942 on January 6, 2014 leading to CME accompanied by a weak flare. We presented the comprehensive study of morphological transformation from simple bipolar configuration to complicated inverse sigmoidal AR, along with its initiation and eruption mechanisms using the multi-wavelength EUV observations from AIA and (vector) magnetograms from HMI. The flare and CME are detected by flare-ribbon evolution and running difference images respectively. 
 
 \begin{figure*}[!ht]
 	\centering
 	\includegraphics[width=.99\textwidth,clip=]{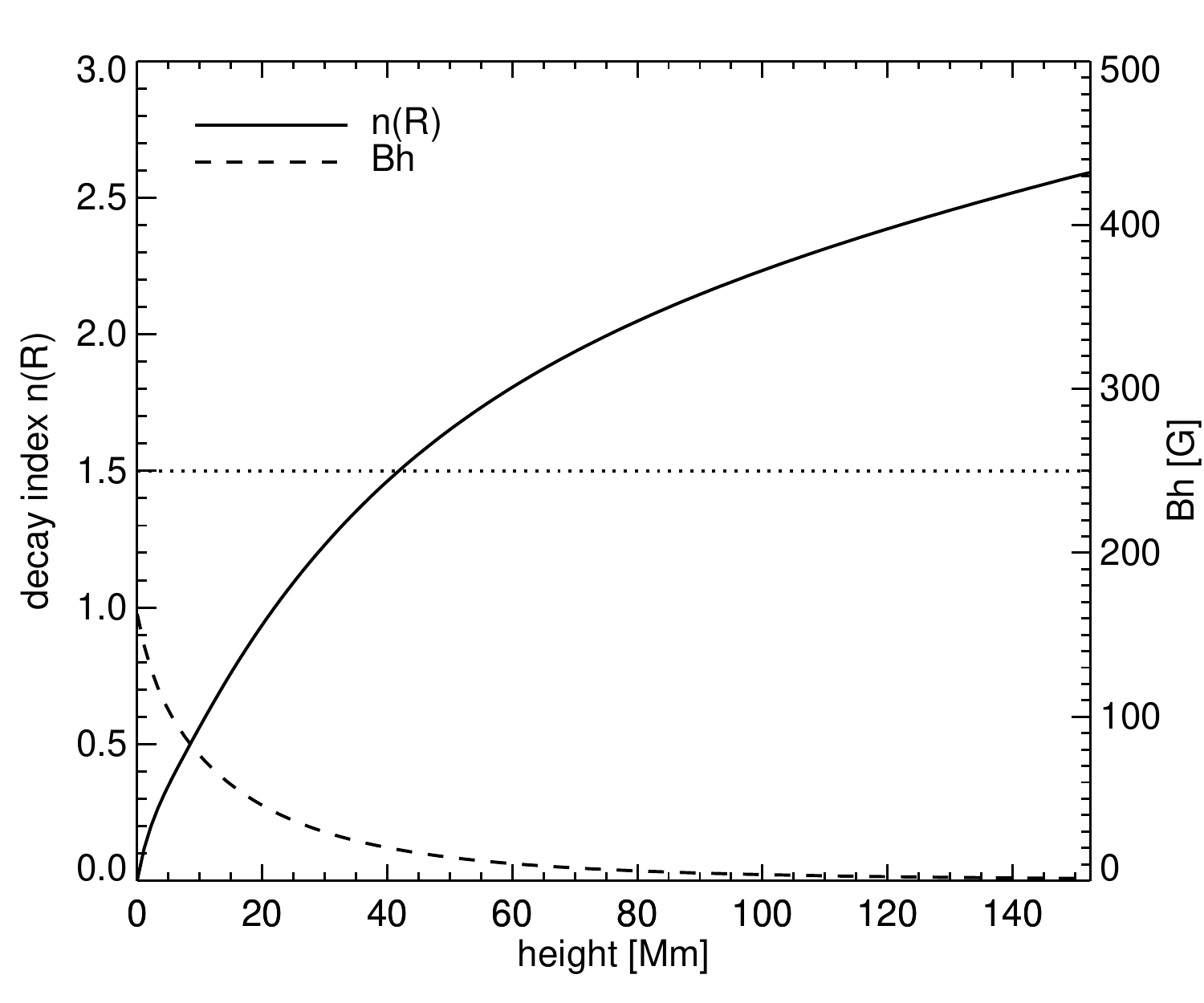}
 	\caption{The decay index $n(z)$ with respect to height. Horizontal magnetic field strength, $B_h$ (dashed curve) is also plotted with y-axis scale on the right. Horizontal dotted line represents critical decay index. } 
 	\label{Fig_di}
 \end{figure*}

Initially the AR is compact and bipolar in nature, and evolved to sheared configuration consisting of inverse-J loops over a couple of days (Figure~\ref{Fig_overall}). Magnetic flux dispersion, shearing and converging motion are observed to play significant role in the development of sigmoidal structure hosting a filament channel as inferred from EUV observations of AIA. Further such evolution leads to flux cancellation and net flux decrease starts early on January 6. Consequently, the filament channel initiated to upward motion from 5:35 UT on January 6, which was suggested for tether-cutting reconnection of inverse J-loops forming continuous inverse S-loops as a flux rope \citep{Moore2001,Amari2003}. While tether-cutting reconnection continues under the converging and shearing motions, the sigmoidal structure losses equilibrium at 21:32 UT leading to CME and weak flare.  \citet{Forbes1991} indicates that as flux cancellation continues near the magnetic neutral line, the flux rope embedded in a bipolar field rises smoothly till it reaches a critical point, at which the flux rope presents a catastrophic behavior, i.e., after the gradual accumulation of magnetic energy, the flux-rope system transits from a stable state to an unstable one containing a current sheet. The catastrophe model of solar eruptions suggests the catastrophic loss of mechanical equilibrium in the magnetic configuration (e.g., \citealt{Forbes2000,Priest2002,Lin2004}). 

Figure~\ref{Fig_cme}c shows when the flux rope rises, overlying loops gets stretched and the inward motion of anti-parallel magnetic field lines lead to the formation of thin current sheet, where the magnetic reconnection sends large amount of the reconnected flux and plasma outward, accounting for the rapid expansion of the ejecta (plasma blobs or plasmoids or CME) and the hot shell surrounding it \citep{Lin2004}. Here the smoothness of the observed eruption is accounted for by the slow reconnection process (local reconnection rate range $ 0.01 -2.14 Vcm^{-1} $) in the current sheet underneath the flux rope. Depending on the amount of reconnected flux, the magnitude of flare emission is visible in X-rays. Lesser the hoop/self force in the flux rope, lesser the coronal disturbance, slower is the CME and its impact. We believe that a small number of inverse J-shaped loops are participated in the tether-cutting reconnection resulting in the weak flux rope and the rest of sheared arcade relaxed in the post phase of the eruption. This is clear from the canceling/converging small flux regions with weak PIL. Accordingly, a mild SXR enhancement is observed due to slow reconnection rate followed by a dominant EUV emission in the late phase of eruption. So, studying such weak events are important to know the connection of photospheric evolution and origin of the Earth-affecting cases.      

Further, \citet{Demoulin2010} and \citet{Kliem2014} showed that the critical conditions for the catastrophic loss of equilibrium also satisfy the torus instability criterion. The rapid decaying of overlying magnetic field with height referred as torus instability and is measured with decay index $n(z) = -\frac{z}{B_h}\frac{\partial B_h}{\partial z}$, where z is the geometrical height from the bottom boundary and $B_h$ is the horizontal field strength. We computed the background field in the entire volume of AR by potential magnetic field approximation. In Figure~\ref{Fig_di}, the $n(z)$ with height in solid curve and $B_h$ in dashed line are plotted. A constant value of $n_{crit}=1.5$ is assumed as critical decay index which corresponds to a critical height of 41.5 Mm. This agrees with the past studies \citep{Cheng2011,Vemareddy2014,Vasantharaju2018}, where the critical heights of eruptive events falls below 42 Mm. 

The magnetic non-potential parameters $\alpha_{av}$, NCN show increasing behaviour with the formation of the sigmoid by slow flux motions for two days prior to the eruption. However, the eruption occurs in a flux cancellation scenario by converging motions from 04:00 UT on January 4. During this period, the free energy exhibits almost steady behavior and $\alpha_{av}$ have a decreasing profile, contrary to the cases of stronger eruptions \citep{Vemareddy2014,Gibb2014}.  It may likely that the the energy release is proceeded in a much long time duration, manifesting the onset of filament rise and eventual eruption driven by converging and canceling flux in the photosphere. 

The eruptions from large ARs are usually associated with one or more clearly visible flux ropes in hot AIA channels. These are associated with sigmoids, filaments/prominences and erupts as CME with a strong X-ray flare \citep{Vemareddy2014,Vemareddy2017, Dhakal2018}. In contrast, the presented case is from an AR where converging/canceling flux region is small with a weak PIL ensuing in a slow reconnection and yet able to produce a successful eruption.

\acknowledgements SDO is a mission of NASA's Living With a Star Program. N.V is a CSIR-SRF, gratefully acknowledges the funding from CSIR-HRDG, New Delhi. P.V. is supported by an INSPIRE grant of AORC scheme under the Department of Science and Technology. We thank the referee for many detailed comments and suggestions.

\bibliographystyle{apj}


\end{document}